\documentstyle[aps,psfig,prb]{revtex}

\begin{document}

\title{ Two-particle pairing and phase separation in a two-dimensional 
Bose-gas with
one or two sorts of bosons
\\
}

\author{ M.Yu.~Kagan  and D.V.~Efremov }
\address{ Max-Planck-Institut f\"ur Physik Komplexer Systeme,
N\"othnitzer Str. 38, 01187 Dresden, Germany  \\
and Kapitza Institute for Physical Problems,Kosygin str. 2,
117334 Moscow, Russia}

\maketitle


\begin{abstract}
We present a phase diagram for a dilute two-dimensional Bose-gas on a
lattice. For one sort of boson we consider a realistic
case of the van der Waals interaction between  particles
with a strong hard-core repulsion $U$ and a van der Waals
attractive tail $V$. For $V< 2 t $, $t$ being a hopping
amplitude, the phase diagram of the system contains  
regions of the usual one-particle Bose-Einstein
condensation (BEC). However for $V>2t$ we have total phase
separation on a Mott-Hubbard Bose solid and a dilute
Bose gas. For two sorts of structureless bosons described by the two band
Hubbard model an $s$-wave pairing of the two bosons of
different sort $\langle b_1 b_2 \rangle \neq 0$ is
possible. The results we obtained should be important for
different Bose systems, including submonolayers of $^4$He,
excitons in semiconductors, Schwinger bosons in magnetic
systems and holons in HTSC. In the HTSC case a possibility
of  two-holon pairing in the slave-bosons theories of
superconductivity can restore a required charge $2e$ of a
Cooper pair.

\end{abstract}


\newcommand{\be}{\begin{equation}}
\newcommand{\ee}{\end{equation}}
\newcommand{\tmu}{\tilde{\mu}}
\newcommand{\eps}{\varepsilon}
\section{Introduction}

It is well known that, in contrast with  two-particle Cooper pairing in
Fermi systems, the essence of a superfluidity in Bose systems is  one particle
Bose-Einstein condensation (BEC). This asymmetry between Fermi and Bose ( two-
particle versus one-particle condensation) was challenged in a pioneering paper
by Valatin and Butler   \cite{Valatin58}. They proposed a BCS-like variational
function for the description of an attractive Bose gas. The most difficult
problem with the validity of their description is connected with a tendency 
toward
phase separation  which arises in attractive Bose systems. Later on Nozieres
and Saint-James \cite{Nozieres82}
conjectured that in  a Bose system with
a short-ranged, hard core repulsion and a van der Waals attractive tail, in
principle, it is possible to create a two-particle bosonic bound
state and to escape collapse. Unfortunately their calculations in a 
three-dimensional  system showed that at least for
one sort of structureless boson either  standard one-particle BEC is more 
energetically beneficial, or that  phase separation takes place earlier 
than the two-particle
condensation. Note that the same result was obtained earlier by Jordanski \cite
{Jordanski65} for the case of weak van der Waals attraction.

The important development of the ideas of  Nozieres and Saint James belongs
to Rice and Wang \cite{Rice88}. The authors claimed that in two-dimensional
(where already an infinitely small attraction leads to the bound state in a
symmetrical potential well) it is possible to realize a two-particle boson
pairing. Moreover, this
two-particle pairing results, for small momenta $p \xi_0 <1$ in a linear, 
soundlike,
dispersion law of quasiparticles at $T=0$ in an analogy with a standard
one-particle Bose-condensation.

To escape a collapse in a 2D attractive Bose gas, the authors of
Ref.\onlinecite{Rice88}
introduced in their model a Hartree-Fock shift of the chemical potential
$\mu_B \sim U n  $, connected with the short-range repulsion $U$.   This shift
in the case of $U>2 V$, where $V$ is the magnitude of the van der Waals tail,
leads to a positive compressibility in the system $\kappa^{-1} = 
d \mu_B / d n = U  - 2V > 0$.

The main goal of a present paper is to construct a phase
diagram of a 2D dilute Bose gas with the van der Waals
interaction between  particles, by taking into account on
equal grounds the full contribution of a hard-core
repulsion $U$ and a van der Waals tail $V$. Throughout the
paper we will consider the lattice model, and will
base our results on the exact solution of the two-particle
$T$-matrix problem presented in  Refs.\onlinecite{Emery90} and 
\onlinecite{Kagan94}. We
will study the possibility of different two-boson pairings,
as well as the possibility of a total phase separation in
the system. We will also consider the two sorts of structureless bosons
described by the two-band Hubbard model. In the case of
attraction between bosons of two different sorts, we will find a
possibility of an $s$-wave two-boson pairing $\langle b_1
b_2  \rangle \neq 0$.

\section{Theoretical model}

The model under consideration  is described by the following Hamiltonian on
the 2D square lattice:
\begin{equation} \label{eq:ham1}
H = -t \sum_{ij}b_{i}^{\dagger}b_j + \frac{U}{2} \sum_{i}
n_i^2 - \frac{V}{2}\sum_{ij}n_i n_j ,
\end{equation}
where $n_i = b_i^{\dagger}b_i$ is a 2D boson density.
We will work in the limit of strong
hard-core repulsion $U \gg \{ V; t\}$, and restrict ourselves mostly to a 
low-density limit $n_B d^2 \ll 1 $, 
$d$ being the interatomic distance. Further on we put
$d=1$.

Note that in the case of $V=0$ model (\ref{eq:ham1}) is just the Bose-Hubbard
model, extensively studied in the literature for the case of 2D $^4$He
submonolayers, as well as for the flux lattices and Josephson arrays in
the type-II superconductors (see Refs.\onlinecite{Fisher89}---\onlinecite{Kampf93}).

We would like to emphasize that  model (\ref{eq:ham1}) is, to some extent, a
Bose analog of the
fermionic $t-J$ model with released constraint, considered by Kagan and
Rice \cite{Kagan94}. After the Fourier transformation  from
Eq. (\ref{eq:ham1}) we obtain:
\begin{eqnarray}
H &=& \sum_{\bf p} \eps_{\bf p}b_{\bf p}^{\dagger}b_{\bf p}
+ \frac{U}{2} \sum_{\bf k_1 k_2 q} b_{\bf k_1}^{\dagger}
b_{\bf k_2}^{\dagger} b_{\bf k_2 -q}  b_{\bf k_1 +q}
\nonumber \\ &-& \sum_{\bf k_1 k_2 q}V({\bf q}) b_{\bf
k_1}^{\dagger} b_{\bf k_1+q} b_{\bf k_2}^{\dagger}b_{\bf
k_2-q},
\end{eqnarray}
where $\varepsilon_{\bf p} = -2 t (\cos{p_x}+\cos{p_y})$ is a bosonic spectrum
on the square lattice and  $V({\bf q})= V (\cos(q_x)+\cos(q_y))$ is a Fourier
transform of the Van der Waals tail. As a result, a total interaction in the
momentum space is given by the formula:
\begin{equation}
\label{eq:Veff} V_{eff}({\bf q}) = \frac{U}{2} -  V({\bf
q}) .
\end{equation}

\section{The $T$-matrix problem.}

An instability toward a two-particle boson pairing manifests itself in the
appearance of a pole at a temperature $T=T_c$ in the solution of the
Bethe-Salpeter equation for the two-particle vertex $\Gamma$ for zero total
momentum of the two bosons $({\bf p}, -{\bf p})$. To proceed to the solution
of this equation, we must solve at first the $T$-matrix problem for the two
particles in vacuum. Here we can use the results of Ref.\onlinecite{Kagan94}, because
the solution of the two-particle problem does not depend upon statistics of
colliding particles. For the $T$-matrix problem it is convenient to expand
$V_{eff}({\bf q})$ in Eq.(\ref{eq:Veff}) in series with the eigenfunctions 
of the irreducible representation of the lattice symmetry group $D_4$. 
This yields
\begin{equation}
\begin{array}{l}
V_{eff}(\mbox{extended $s$-wave}) = \frac{U}{2} -
\frac{V}{2}(\cos{p_x}+ \cos{p_y}) (\cos{p_x'}+\cos{p_y'}),
\\ V_{eff}(\mbox{$p$-wave}) =  - \frac{V}{2}(\sin{p_x}\sin{p_x'} +
\sin{p_y}\sin{p_y'}), \\ V_{eff}(\mbox{$d_{x^2-y^2}$}) = -
\frac{V}{2}(\cos{p_x}- \cos{p_y}) (\cos{p_x'}-\cos{p_y'}),
\\
\end{array}
\label{eq:Vchan}
\end{equation}
 where ${\bf q} = {\bf p} - {\bf p'}$ is a transfered momentum.
Note that, for spinless bosons, which we formally consider in Eq. (1), 
the total spin
of the Bose pair is zero. Hence only  $s$- and a $d$-wave pairings are
allowed by the symmetry of the pair $\Psi$ function. A $p$-wave pairing is allowed only for an odd total
spin of the two bosons. Nevertheless  we will conserve the results for
the $p$-wave pairing in our paper because the generalization of Eq.(1) for the
case of bosons with internal degrees of freedom is straightforward.
The $T$-matrix
problems  for $p$- and $d$-wave channels are very simple. Solutions of
these problems for the two particles with a total momentum zero and a total energy $E$
yield

\begin{equation} \label{eq:Tmat}
T_{d;p} (E) = -\frac{ \frac{1}{2} V}{\left( 1+ \frac{1}{2}V
I_{d;p} \right)},
\end{equation}
where
\begin{equation} \label{eq:Idp}
I_{d;p} = \int\limits_{0}^{2 \pi} \int\limits_{0}^{2 \pi} \frac{d p_x ~ d p_y}{(2 \pi)^2}
\frac{|\phi_{d;p}|^2}{E+4 t (\cos p_x + \cos p_y)},
\end{equation}
and the functions $\phi_d = \cos{p_x} - \cos{p_y}$ and
$\phi_p = \sin{p_x}+ i \sin{p_y}$
correspond, respectively, to the $d$- and  $p$-wave channels.

Let us find the thresholds for the bound states in the $d$- and the $p$-wave
channels. The  appearance of a bound state means that $E=-W-|\tilde{E}|$,
where $W = 8t$ is a bandwidth. For the threshold $\tilde{E} = 0$. An exact
solution of Eqs.  (\ref{eq:Tmat}) and (\ref{eq:Idp}), which involves the
calculation of elliptic integrals of first and second order, yields
\begin{equation}\label{eq:7}
\begin{array}{l}
\displaystyle \left( \frac{V_c}{4 t}
\right)_{p-\mbox{\it wave}} = \frac{1}{\displaystyle
1-\frac{2}{\pi}} \approx 2.8 , \\ \displaystyle \left(
\frac{V_c}{4 t} \right)_{d-\mbox{\it wave}} =
\frac{1}{\displaystyle \frac{4}{\pi}-1} \approx 3.7 .
\end{array}
\end{equation}
Note that a threshold for a $p$-wave pairing is lower.
Now let us proceed to an $s$-wave channel. Here an ordinary
$s$-wave pairing is suppressed by large hard-core repulsion $U$, however an
extended $s$-wave pairing with a symmetry of the order parameter
$\Delta_s = \Delta_0 (\cos p_x + \cos p_y)$ is allowed. In real space this
pairing corresponds to the particles on the neighboring sites. Moreover the pair
$\Psi$  function is zero in the region of a hard core ($r<r_0$) and is
centered (has  a maximum) in the region of a van der Waals attraction for $r
\sim r_1$ (see Fig. 1). On the lattice $r_0\sim d/2$ and $r_1 \sim d$.

One can see that the $\Psi$ function has a region of zero value zero. But it has no nodes because 
it does
not change its sign for all values of $r$. The rigorous calculation of the
threshold for an extended $s$-wave pairing gives \cite{Emery90,Kagan94}
\begin{equation}
\label{eq:swave} \left( \frac{V_c}{4t}
\right)_{s-\mbox{\it wave}}  =1.
\end{equation}
Moreover for $V>V_c$ an energy of the bound state has the form
\begin{equation}\label{eq:9}
|E_b| = |\tilde{E}| = 8 W \exp
\left\{
-\frac{\pi V }{(V-V_{crit})}
\right\}.
\end{equation}
Of course, in a strong coupling case $V \gg W$ 
$$
|E_b| \approx V.
$$
The $T$ matrix in an  $s$-wave channel for small and intermediate values  of
$V$ is given by
\begin{equation} \label{eq:Ts}
T_s(|\tilde{E}|) = \frac{ \displaystyle W \left(1-\frac{V}{4 t} \right)}
{\displaystyle \frac{1}{\pi} \left(1-\frac{V}{4 t} \right)\ln \frac{8 W}
{|\tilde{E}|}  - \frac{V}{4t}}.
\end{equation}
The most important is that a strong Hubbard repulsion $U$ acts only as an
excluded volume and effectively drops out from Eq. (\ref{eq:Ts}) at low
energies.
For $V \ll 4t$ the $T$ matrix
$$
T_s(|\tilde{E}|) \approx \frac{\pi W}{\ln \frac{8 W}{|\tilde{E}|}}
$$
corresponds to repulsion and coincides with the $T$ matrix for the 2D
Bose-Hubbard model at low density. Of course, at very high energies
the $T$ matrix contains an additional pole  $E \approx U$  which corresponds
to an antibound state in a total analogy with the fermionic Hubbard model.
For $V=4t$: $T_s(|\tilde{E}|) =0$ and there is no interaction at all.
Finally, for $V>4t $, $T_s(|\tilde{E}|)<0$ corresponds to an attraction and
reflects the 
appearance of the bound state.

\section{Bethe-Salpeter equation for an $s$-wave pairing of the two bosons}

Let us consider at first the most interesting case of $V>4t$ and find the
critical temperature for an extended $s$-wave pairing of the two bosons.
The solution of the Bethe-Salpeter equation for bosonic  systems reads \cite{Griffin98}
\begin{equation} \label{eq:11}
\Gamma_s = \frac{T_s}
{ \displaystyle 1+T_s \int \int \frac{d p_x d p_y}{(2 \pi)^2} \frac{\coth\frac{\varepsilon_{\bf
p}-\mu}{2 T}}
{2 (\varepsilon_{\bf p}-\mu)}}.
\end{equation}

For a low density of bosons $n_B \ll 1$ one has
$$
\varepsilon_{\bf p} \approx - 4 t + \frac{p^2}{2 m},
$$
where $m = 1/2t$ is a boson mass.
Accordingly $\mu = - 4t + \tilde{\mu}$ and $\xi_{\bf p} = \varepsilon_{\bf
p}-\mu = p^2 / 2m-\tilde{\mu}.$

The most substantial difference of Eq. (\ref{eq:11}) from an analogous
fermionic equation is the replacement of  $\tanh{\xi / 2T}$ by
$\coth{\xi /2T}$ in its kernel.
Moreover, as shown in Ref.\onlinecite{Miyake83} for the 2D Fermi gas
$\tilde{\mu}=\varepsilon_F - |E_b| / 2$. So, in a weak-coupling case, when
$\varepsilon_F \gg |E_b|$, the chemical potential $\tilde{\mu}\approx \varepsilon_F>0$ is
positive. In contrast to this we shall see below  that a bosonic chemical
potential $\tilde{\mu}$ is always negative even in the weak-coupling case, when a binding
energy is much smaller than a degeneracy temperature:
$$
|E_b| \ll T_0 = \frac{2 \pi n }{m }.
$$
Another very important point is that the $T$ matrix, which enters into the
Bethe-Salpeter equation, must be calculated for a total energy $\tilde{E}=2
\tilde{\mu}$ of colliding bosons. The chemical potential $\tilde{\mu}$ can be
determined from the requirement of the number of particle conservation.
This requirement yields 
\begin{equation} \label{eq:n_b}
n_B = \int \int \frac{d^2 p}{(2 \pi)^2} \frac{1}{
\exp \left\{\frac{ p^2 /2m-\tilde{\mu}}{T}  \right\} -1}.
\end{equation}

From Eq.(\ref{eq:n_b}) for the temperatures $|E_b|<T<T_0<W $ we obtain
\begin{equation}
\tilde{\mu} = - T \exp \left(
-\frac{T_0}{T}
\right)<0 .
\end{equation}
Note that a standard  Hartree-Fock shift $n U$ drops out from the 
expression for $\xi_{\bf p}
=\eps_{\bf p }-\mu$ both in the Bethe-Salpeter equation (\ref{eq:11}) and in
the equation for the number of particle conservation (\ref{eq:n_b}).
Now we are ready to solve the Bethe-Salpeter equation (\ref{eq:11}). The critical
temperature $T_c$ corresponds to the pole in Eq. (\ref{eq:11})

\begin{equation} \label{eq:15}
1+ \frac{m T_s(2 \tilde{\mu})}{2 \pi} I =0 ,
\end{equation}
where
\begin{equation} \label{eq:16}
I = \int\limits^{\sim W/T_c}_0 d y ~~\frac{\coth \left( \displaystyle
y+ \frac{|\tilde{\mu}|}{2T_c} \right) }{\displaystyle y + \frac{|\tilde{\mu}|}{2T_c}},
\end{equation}
and $$y= \frac{p^2}{4 m T_c}.$$

An analysis of Eq.(\ref{eq:16}) shows that the main contribution to the integral
comes from the lower limit of integration.

Hence providing $|\tilde{\mu}|/T_c \ll 1$  we have
\begin{equation}
I \approx \int\limits^{W/T_c}_{0} \frac{d y}{\left( \displaystyle y +
\frac{|\tilde{\mu}|}{2 T_c}\right)^2} \approx \frac{2 T_c}{|\tilde{\mu}|}.
\end{equation}
As a result Eq. (\ref{eq:15}) can be represented in the following form:
\begin{equation}\label{eq:18}
\frac{T_c}{|\tilde{\mu}|} = - \frac{\pi}{m T_s(2\tilde{\mu})}.
\end{equation}
It is useful now to represent $T_s(2\tilde{\mu})$ in terms of the binding
energy $E_b$.
Utilizing Eqs. (\ref{eq:9}) and (\ref{eq:Ts}) we can write
\begin{equation}\label{eq:18}
T_s(2\tilde{\mu}) = - \frac{\pi W}{\ln \frac{2 |\tmu|}{|E_b|}} =
-\frac{4 \pi}{m \ln\frac{2 \tilde{|\mu|}}{|E_b|}}
\end{equation}

It is important to mention here that $\tilde{\mu}<0$, and
hence the $T$ matrix in Eq. (\ref{eq:18}) does not contain an imaginary part.
In the fermionic case $\tilde{\mu} = \varepsilon_F >0$,  and the $T$ matrix
contains an imaginary part  corresponding to the resonant scattering.
 As a result, from Eq. 
(\ref{eq:18}) we obtain:
\begin{equation} \label{eq:20}
4 \ln \frac{2 |\tilde{\mu}|}{E_b} = \frac{|\tilde{\mu}|}{T_c}.
\end{equation}

Assuming that $|E_b| \ll T_c \ll T_0$ , we obtain
$$
\tilde{\mu}(T_c) = - T_c \exp\left(
-\frac{T_0}{T_c}
\right) 
$$
and
$$
\frac{|\tilde{\mu}|}{T_c} = \exp \left\{ - \frac{T_0}{T_c} \right\}.
$$
Later on  we will justify this assumption.

As a result  from Eq. (\ref{eq:20}) we will obtain
\begin{equation}\label{eq:Tc}
T_c = \frac{T_0}{\ln \left(
\frac{1}{4}\ln \frac{2|\tilde{\mu}|}{|E_b|}
\right)}.
\end{equation}

Recall that  in the case of the fermionic pairing in two dimensionsa critical temperature
reads \cite{Miyake83}
 $$T_c = \sqrt{2 \varepsilon_F |E_b|} . $$

Let us analyze expression (\ref{eq:Tc}). As we already know
$$
|E_b| = 8 W \exp\left\{
-\frac{1}{\lambda}
\right\},
$$
where
\begin{equation}
\lambda = \frac{(V-V_{cs})}{\pi V}.
\end{equation}
Then a condition $|E_b| \ll T_0$ means
\begin{equation}
\lambda \ll \frac{1}
{\ln \frac{W}{T_0}}
 \ll 1 .
\end{equation}

Hence $\ln T_0 / |E_b|= 1 / \lambda- \ln W / T_0 \approx
1 / \lambda$, and
\begin{equation}\label{eq:24}
T_c \approx \frac{T_0}{\ln \frac{1}{4}\ln\frac{T_0}{|E_b|}}
\approx \frac{T_0}{\ln \frac{1}{4 \lambda}},
\end{equation}
 which is in an agreement with Ref.\onlinecite{Rice88}.
Note that $T_c$ from Eq. (\ref{eq:24}) satisfies the conditions $|E_b| \ll T_c \ll
T_0$, so an assumption used for the derivation of  $T_c$ is justified.

For $T<T_c$ the spectrum of the quasiparticles acquires a gap:

\begin{equation}
E_{\bf p} = \sqrt{(\varepsilon_{\bf p}+|\tilde{\mu}|)^2-\Delta^2}.
\end{equation}
Note that at low densities of bosons a gap $\Delta$ becomes isotropic
in the principal approximation.

The gap $\Delta$ together  with the chemical potential
$\tilde{\mu}$ must be defined self-consistently from the
two coupled equations
 \be \label{eq:26} 1=
\frac{\lambda}{4}
\int\limits_{\frac{|\tilde{\mu}|}{2T}}^{\sim \frac{W}{2T}}
d z \frac{\coth \sqrt{z^2 - \frac{\Delta^2}{4
T^2}}}{\sqrt{z^2 -\frac{\Delta^2}{4T^2}}} , \ee

\be \label{eq:27}
n_B =  \frac{\lambda}{4} \int\limits_{|\tilde{\mu}|}^{\sim W}
d \xi \frac{1}{\exp\left\{\frac{\sqrt{\xi^2 - \Delta^2}}{T}\right\}-1} ,
\ee
where $\xi = \varepsilon + |\tilde{\mu}|$ and  $z = \xi / 2T$.

Of course, the solution of the system of equations Eqs. (\ref{eq:26}) and  (\ref{eq:27})
exists only if $|\tmu|\geq \Delta$, or, in other words, only if $E_g^2 =
|\tmu|^2-\Delta^2>0$. The exact solution of these equations yields
for zero temperature in an agreement with \cite{Rice88}
\be
|\tmu(T=0)|= \Delta = \frac{|E_b|}{2}. \ee This result is
very important. It justifies our scenario, leading to a
linear, soundlike spectrum of the quasiparticles for a small
momenta $p$.
Indeed
 \be \label{eq:29} E_{\bf p}
=\sqrt{\varepsilon_{\bf p}^2 + 2  \varepsilon_{\bf
p}|\tmu|} = \sqrt{\varepsilon_{\bf p}^2 + \varepsilon_{\bf
p}|E_b |} . \ee
From Eq. (\ref{eq:29}) for the case $p \xi_0
\ll 1$, where $\xi_0 = 1 / \sqrt{2 m |E_b| }$ is the
coherence length of the boson pair, we immediately obtain a
linear  dispersion law: \be \label{eq:30} E_p = c p . \ee
In (\ref{eq:30}) $c^2= |E_b|/2 m$ is a sound
velocity squared. This means that an inverse compressibility
of the system $\kappa^{-1}=c^2$ is positive. This fact proves
the stability of a superfluid paired state and excludes the possibility
of the collapse of the pairs in the system. Note also that close
to $T_c$ one has
\be
\Delta(T) \approx \Delta (0) \sqrt{\frac{T_c-T}{T_c}}, \ee
which is similar to  the BCS theory. We would like to
mention that bosonic pairs in the
limit $|E_b| \ll T_0$ are extended in  full analogy with the
BCS theory. That is  the coherence length in this limit,
\be
\xi_0 \gg \frac{1}{\sqrt{n}} \gg 1 , 
\ee
is larger than the mean distance between the bosons. The Bose pairs are
strongly overlapping in this limit. The pairing takes place in the momentum space
in an analogy with the Cooper pairing in the BCS picture of superconductivity.

In the opposite limit $|E_b| \gg T_0 $, the situation  closely resembles the
bipolaronic limit for the fermionic systems \cite{Alexandrov81,Nozieres85}.
That is, the creation of the bosonic bound pairs is associated with the
crossover temperature \cite{Kagan00}
\be \label{eq:32}
T^{*}  = \frac{|E_b|}{\ln( 1 /n)} .
\ee
The Bose condensation of the pairs  occurs at lower temperature
\cite{Fisher88,Popov72}:
\be \label{eq:33}
T_c = \frac{T_0}{\ln \ln (1/n)} .
\ee
Note that this temperature is obtained from the condition
$\mu_b(T_c) = - T_c \exp{\left(-T_0 / T_c\right)}+f_0 T_c =
0$, where
$f_0 = 1 / \ln (1/n)$ is a repulsive interaction between the local pairs.
So the superfluid transition takes place only for a residual repulsion between
the pairs. Also note that in a dilute Bose gas  the Berezinski-Kosterlitz-Thouless
contribution of vortices \cite{Kosterlitz73} is important only very close to
$T_c$, so the mean
field expression (\ref{eq:33}) gives a very good estimate for the exact  BKT
critical temperature:
$$
(T_c - T_{BKT})/T_c \sim 1/ \ln \ln (1/n) \ll 1.
$$
In the case of the local pairs the coherence length is small:
\be
\xi_0  \ll \frac{1}{\sqrt{n}} .
\ee
The pairs are compact, and the pairing takes place in the real space.

\section{Possibility of  $p$-wave and  $d$-wave pairing of the two bosons}

Now let us analyze the solution of the Bethe-Salpeter equation for  $p$-
and  $d$-wave boson
pairings. Here the critical temperatures should be found from the conditions
\be
1+ T_{p;d}(2 \tmu)\tilde{I}_{p;d} =0,
\ee
where
\be
\tilde{I}_{p;d} = \int\limits_{0}^{2 \pi}\int\limits_{0}^{2 \pi}
\frac{d p_x d p_y}{(2\pi)^2} \frac{\coth{\frac{\varepsilon - \mu}{2T_c}}
}{2(\varepsilon - \mu)} |\phi_{p;d}|^2 .
\ee
In a low-density limit the $\phi$ functions can be approximated by the following
expressions:
$$
\phi_p \approx p_x + i p_y = p e^{i \phi},
$$
$$
\phi_d \approx \frac{1}{2} ( p_x^2 -  p_y^2) = \frac{1}{2} p^2 \cos{2 \phi}.
$$
Hence after an angular integration we obtain:
\be
\begin{array}{l} \displaystyle
\tilde{I}_p = \frac{m}{2 \pi }\int p d p\frac{\coth \frac{\xi}{2T_c}}{2
\xi}  p^2 , \\ \displaystyle
\tilde{I}_d = \frac{m}{16 \pi }\int p d p \frac{\coth \frac{\xi}{2T_c}}{2
\xi}  p^4 ,
\end{array}
\ee
where again $\xi = p^2 / 2m +|\tmu|$.

Additional factors $p^2$ and $p^4$ in the integral expressions for $\tilde{I}_p$
and  $\tilde{I}_d$ reflect a well-known fact, that for slow 2D particles in
vacuum an $s$-wave harmonics of the scattering amplitude behaves as $f_0 \sim
\ln 1/ p^2$, whereas for a magnetic number $m\neq 0$, the scattering amplitude
vanishes for $p$ goes to zero as $f_m \sim p^{2 m}$. The additional factor $p^4$
leads to the absence of an infra-red singularity for $\varepsilon \to 0 $ in
$\tilde{I}_d$:
\be
\tilde{I}_d \sim \int \frac{d \eps~ \eps^2}{\eps^2} \sim \eps \to 0.
\ee

For the $p$-wave channel the infra-red singularity becomes logarithmically
weak:
\be
\tilde{I}_p \sim \int \frac{d \eps ~\eps }{\eps^2}  \sim \ln \eps.
\ee
This means that the Bethe-Salpeter equation has no solutions in  $p$- and
$d$-wave
channels for $|V| / t<1$.

Hence the boson pairing with a large coherence length
$\xi_0 > 1 / \sqrt{n}$ is absent in a $p$-wave channel as well 
as in a $d$-wave channel. Here
only the limit of the local pairs is possible. For  $p$- and  $d$- wave
channels  
local pairs are created at the crossover temperatures given by Eq. (\ref{eq:32}).
In this case the  binding energies for  $p$- and  $d$-wave channels   read, correspondingly,
\be \label{eq:40}
|E_b^{p}| \approx W \frac{(V-V_{cp})}{V} \frac{1}{\ln \frac{V}{(V-V_{cp})}},
\ee
\be \label{eq:41}
|E_b^{d}| \approx W \frac{(V-V_{cd})}{V} .
\ee
Note that $V_{cp}$ and $V_{cd}$ in Eqs. (\ref{eq:40}) and (\ref{eq:41}) are the
thresholds for the
$p$- and $d$-wave pairings given by Eq. (\ref{eq:7}). We would also like to
mention 
 that for a fixed $V$:
$|E_b^{p}|>|E_b^{d}|$. Providing that the interaction between the local pairs
is repulsive, the temperature of the Bose condensation of the local pairs in
these two cases is again given by Eq. (\ref{eq:33}).
Summarizing this case we see that,
while the crossover temperatures $T^{*}$ are different for the $s$-,  $p$- and
$d$-wave local pairs, their critical temperatures
coincide with each other in a principal order.

\section{Total phase separation}
As we discussed in the Sec. 1, the real collapse is
prohibited in our system by large Hubbard repulsion $U$.
However the phase separation on the two large clusters is
allowed. The first cluster corresponds to the Mott-Hubbard
Bose solid. In this cluster $n_b \to 1$, that is,  each site
is practically occupied by one boson. Such a cluster is
localized due to  Mott-Hubbard considerations. It has no
kinetic energy. However it has a potential energy of the
order of $-2 V$ for one particle. A second cluster has a very
small boson density $n \to 0$. In this cluster for $V<4 t$ the energy
per particle is $-W/2 + f_0 n $, where $f_0 = m / 4 \pi \ln 1/n$
is a two-particle $T$ matrix in the absence of a bound state.
Rigorously speaking, at
a given bosonic density $n$ the phase separation results in
the formation of the two clusters with the densities $n_1
> n$  and $ n_2 <n$, where $n_1$ is close to or identically equal to 1.
The phase
separation for $V<4 t$ takes place if the energy per particle in the
cluster with the density $n_1$ becomes smaller than the
energy per particle in the cluster with the density $n_2$
\be \label{eq:43}
 -2 \eta V \leq -W/2  ,
 \ee
where
$\eta$ is an unknown numerical coefficient of the order of $1$.
Note that in
the fermionic $t-J$ model, considered in Ref.\onlinecite{Kagan94},
the Mott-Hubbard cluster with $n=1$ has an antiferromagnetic order. Hence
instead of $\eta V$ in (\ref{eq:43}) one should write $1.18
J$ -- the energy per spin of the 2D antiferromagnetic. As a result, in a
fermionic case $J_{ps}=3.8t$. In our system $V_{ps} \approx 2 t$,
due to the absence of  kinetic energy and zero point energy
in the case of structureless  bosons.
In the same time for $n \to
1$ the phase separation between the Bose solid and the
one-particle BEC takes place. According to
Ref.\onlinecite{Dagotto93}
for $n_B \rightarrow 1$, the phase separation takes place already
for small values of $|V|/t$.

In principle, another scenario of the phase separation connected
with the creation of quartets \cite{Nozieres98} is also possible in our
system. It requires an evaluation of the four-particle vertex which is
impossible to do analytically.  However we think that our scenario of total phase
separation takes place for smaller values of $V/t$ than the quartet
formation. This is in agreement with numerical calculations
\cite{Dagotto93} for the 2D fermionic $t-J$ model.

\section{Phase diagram of the system}
In this section we will complete the phase diagram of the system.
At first, note that for $V<V_{ps}$ (when the $T$ matrix for an $s$-wave channel
is repulsive) we have at low density a standard Bose gas with a hard-core
repulsion. It will be unstable toward  a standard one-particle BEC  at a
critical temperature given again by Eq. (\ref{eq:33}). For
$V>2t$ a total phase separation on two large clusters takes
place in our system. One of these clusters contains  a Mott-Hubbard
Bose solid, another one contains a Bose gas with
one particle condensation.

For large densities $n=n_c \leq 1$ ($n_c \equiv 1$ in
Ref.\onlinecite{Kampf93} for structureless bosons) the system
will undergo a transition to the
Mott-Hubbard Bose solid.
As a result, 
on a qualitative level the phase diagram for our system has the form,
presented in Fig.2.
Note that our phase diagram should be  important for
the understanding of the physics of the gas of kinks and steps on a solid
interface of $^4$He. It will
be interesting in this context to obtain phase separation in a system of kinks
on a quantum atomically rough surface of $^4$He. Another
possible application of our results  can be connected with
the study of biexcitonic pairing in semiconductors \cite{Keldysh92}. In
this context we must mention an interesting recent paper of
Lozovik {\it et al }. \cite{Lozovik99}.

\section{Two-band Hubbard model for the two sorts of bosons}

Let us consider the two-band Hubbard model for the two sorts of
structureless bosons. The Hamiltonian of the system has the
form
\begin{eqnarray}
H =&& -t_a \sum_{ij} a_{i}^{\dagger}a_j -t_b
\sum_{ij} b_{i}^{\dagger}b_j 
  \nonumber \\
&&
 + \frac{U_{aa}}{2} \sum_{i} n_{i a}^2 + \frac{U_{bb}}{2} \sum_{i} n_{i b}^2
- \frac{U_{ab}}{2}\sum_{i}n_{ia} n_{ib} ,
\end{eqnarray}
where $t_a$ and $t_b$ and $n_a$ and $n_b$ are, respectively,
the hopping matrix elements
and densities  for bosons of sorts $a$ and $b$. For
simplicity we will consider the case $t_a=t_b$,
corresponding to the equal masses $m_a = 1/(2t_a) = m_b$. We
also assume that the bottoms of the bands coincide.
In the Hamiltonian $U_{aa}$ and $U_{bb}$ are Hubbard onsite
repulsions for bosons of sorts $a$ and $b$. Finally
$U_{ab}$ is an onsite attraction between bosons of two
different sorts.

Let us consider the low-density limit, when both $n_a \ll 1$
and $n_b \ll 1$. In this limit we must replace the Hubbard
interaction  $U_{ab}$ by the
corresponding $T$ matrix.
The relevant expression for the $T$ matrix $T_{ab}$ is given by
\begin{equation}
T_{ab}(\tilde{E}) =
\frac{ U_{ab}}{1 - U_{ab} \int \frac{d^2 p }{ 4 \pi^2 }
\frac{1}{ p^2/m  +|\tilde{E}| } } ,
\end{equation}
where $\tilde{E}$ is given again by $\tilde{E} = E + W$. The
$T$ matrix has the pole for the energy
\be
|\tilde{E}|= |E_b| =  W \exp \left\{
-\frac{4 \pi}{m U_{ab}}
\right\}.
\ee
In the extremely strong coupling case $U_{ab} > W$ the pole
corresponds to the energy $|E|=|E_b| = U_{ab}$.
The pole in the $T$ matrix reflects the appearance of the bound state
of the two bosons of different sorts.

Now we can solve the two-particle problem in the presence of
the bosonic background. A simple analysis shows that only
local bosonic pairs (bipolarons) are possible in our case. They are
formed at a high crossover temperature $T^* = |E_b| / \ln (W/T_0)$,
 where $T_0 = \min \{T_{0}^a, T_{0}^b\}$, and  $T_0^{a}$ and $T_0^{b}$ are degeneracy
 temperatures for bosons of the two sorts. Correspondingly the pairs are bose
 condensed at lower temperature $T_c^{ab} = T_0/ \ln (
 1/\lambda)$, where  $\lambda = m U_{ab}/(4
 \pi)$. Our results are
 valid is the case $|E_{b}|> \left\{ T_0^{a}, T_0^{b}\right\}
 $. In the opposite case of higher densities, when at least
 one of the temperatures $T_0^{a} $ or $T_0^{b}$ is larger
 then $|E_b|$, we have at first a standard one particle
 condensation for bosons with higher density. As a
 result the two-particle pairing between bosons of
 different sort can take place only as a second superfluid transition
 inside the superfluid phase with one particle BEC.
Let us now analyze the stability of our system with respect
to collapse.
For simplicity we
consider  an extremely strong coupling case $\{U_{aa},U_{bb}, U_{ab}\} >
W$. In this case the local pairs of two bosons of different
sorts have onsite character $\langle a_i b_i\rangle \neq 0$.
To escape collapse in this case we must satisfy the following
stability criteria: namely, we must prevent
the collapse of  $\langle a b\rangle$ pairs on one
site (a creation of the quartets  $\langle a a  b
b\rangle$). This requires the inequality $U_{aa}+U_{bb} - 4
U_{ab}>0$.
These results can be important for slave-boson and
Schwinger-boson theories.

\section{Slave-boson formulation of the {\normalsize $t-J$} model.\\  Application to HTSC.}

We conclude our paper with a brief discussion of the problems which arise in
a slave-boson formulation of the $t-J$ model. In this formulation close to
half-filling ($n \to 1$)  an  electron is
represented as a product of holon and spinon \cite{Anderson87}:
$$
c_{i \sigma}^{\dagger}  = f_{i \sigma}^{\dagger} b_i .
$$

A superconductive $d$-wave gap
$$\Delta_d = \langle c_{i \uparrow} c_{j \downarrow} -c_{i \downarrow} c_{j
\uparrow} \rangle  $$
 is a direct product $$\Delta_d = \Delta_{sp}\Delta_{h}^{\dagger}$$ of a spinon
$d$-wave
gap  $$\Delta_{sp} =  \langle f_{i \uparrow} f_{j \downarrow} -f_{i
\downarrow} f_{j \uparrow}  \rangle $$
and a holon $s$-wave gap $$\Delta_{h} =  \langle b_i b_j \rangle. $$
Then a natural question arises whether $ \langle b_i \rangle \neq 0$ and, accordingly,
$\Delta_h = \langle b_i \rangle   \langle b_j \rangle $, or $ \langle b_i
\rangle  = 0$ but   $ \langle b_i b_j \rangle  \neq 0$. In other
words, whether a one particle or  two-particle condensations of the holons takes place
in our system \cite{Wen98,Kotliar88}.

This problem is a very difficult one and, surely deserves  a very careful
analysis. Our preliminary considerations show, however, that the more beneficial
conditions for the two-particle condensation may arise in the SU(2) formulation
of the
$t-J$ model \cite{Lee98,Salk00}. In the standard U(1) formulation of the model
\cite{Lee92} an effective potential of the two-holon interaction on neighboring
sites appearing after the Hubbard-Stratonovich transformation has a form
\be
\left(
\frac{8 t^2}{J} - \frac{J}{4}
\right) \sum_{ij} b_i^{\dagger } b_j^{\dagger } b_i b_j ,
\ee
and thus corresponds to the repulsion for $t>J$. This observation excludes the
possibility of the two-holon pairing in the U(1) formulation of the $t-J$ model.

In the SU(2) case it will be desirable to derive conditions when $ \langle b_1
\rangle =
\langle b_2 \rangle = 0$  but $ \langle b_1 b_2 \rangle \neq 0$. For
such a nondiagonal pairing, as  already discussed above 
it is easier to satisfy the stability criteria \cite{Nozieres82}.  Note also that
the same situation with two sorts of bosons and a possible
attraction
between them can be realized for 2D magnetic systems. We obtain the
corresponding bosonic Hamiltonian here after a Schwinger
transformation of spins \cite{Arovas88,Irhin99} in extended Heisenberg
models.
Work along these lines is now in progress.

\section{Conclusions}
In conclusion we analyzed the possibility of the formation
of boson pairs with $s$-wave symmetry and an appearance of 
 total phase separation in a 2D
Bose gas. In addition we considered the case of
boson pairs with  symmetry of $p$- and $d$-wave
type. We also constructed the qualitative phase diagram for
the 2D Bose gas with the van der Waals interaction between
the particles, which, besides a standard one particle BEC,
contains  regions of the Mott-Hubbard Bose solid and a
total phase separation. We also consider  the situation for
two sorts of bosons described by the two-band Hubbard model, and
found the conditions for the two-particle pairing between
bosons of different sorts. We discuss the applicability of
our results for the different physical systems ranging from
submonolayers of $^4$He and excitons in the semiconductors
till Schwinger bosons in magnetic systems and holons in the slave-boson 
theories of HTSC.

The authors acknowledge helpful discussions with  P. Fulde,
O.~Fischer,  Yu. Kagan, N.M.~Plakida,
G. Khaliullin, S.H.S.~Salk, M.A. Baranov, D. Ivanov, G.~Jackeli,
G.V. Shlyapnikov and V.I.~Marchenko. This work was also supported by Russian 
Foundation for Basic Research (grant 02-02-17520), Russian President 
Program for Science Support (grant  96-15-9694) and by Grant of 
Russian Academy of Sciences for young  scientists.

\begin{figure}[t]
\setlength{\unitlength}{0.0001\columnwidth}%
\begin{picture}(0,0)(0,-400)%
\psfig{file=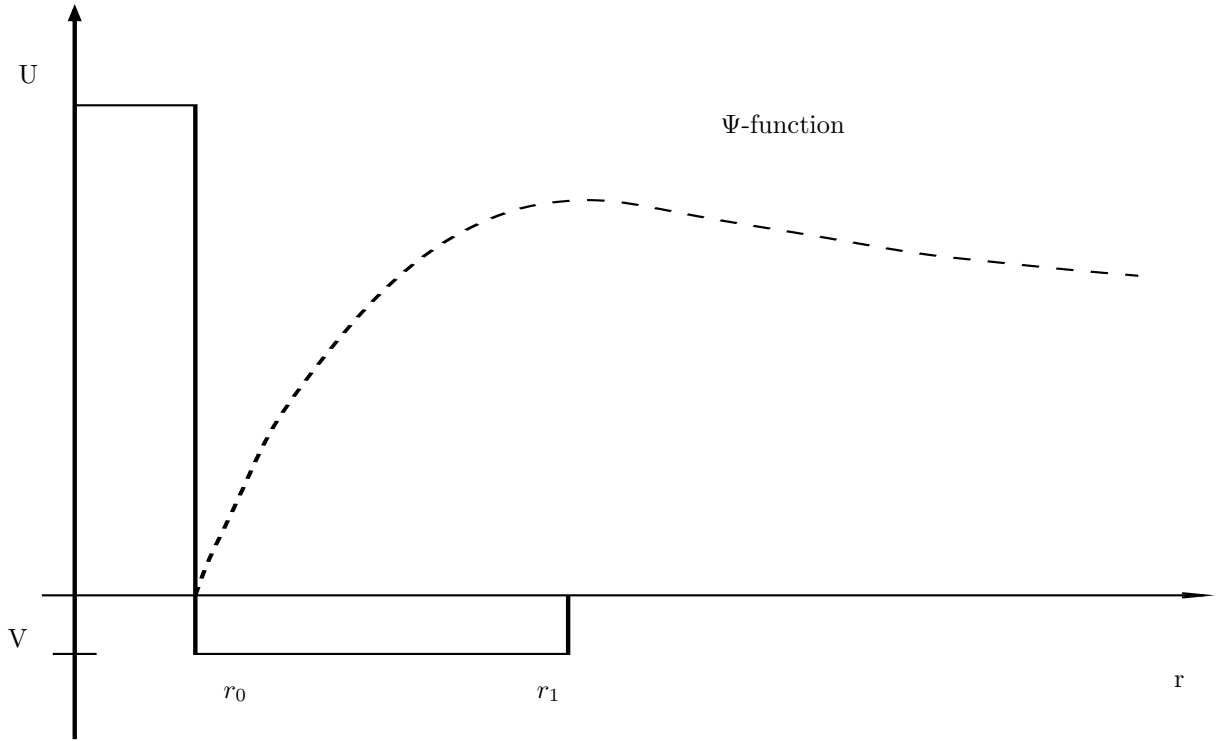,width=0.9\columnwidth,height=0.55\columnwidth}%
\end{picture}%
\begin{picture}(8283,5816)(2176,-8569)
\put(2176,-7511){\makebox(0,0)[lb]{V}}
\put(7126,-3936){\makebox(0,0)[lb]{$\Psi$-function}}
\put(5851,-7861){\makebox(0,0)[lb]{$r_1$}}
\put(3676,-7861){\makebox(0,0)[lb]{$r_0$}}
\put(10276,-7786){\makebox(0,0)[lb]{r}}
\put(2251,-3586){\makebox(0,0)[lb]{U }}
\end{picture}
\caption{The $\Psi$-function of an extended $s$-wave pairing. $r_0$ is a radius
of a hard core repulsion, $r_1$ is a radius of a Van der Waals attraction. On the
lattice $r_0 \sim d/2$ and $r_1 \sim d$}
\label{fig:1}
\end{figure}

\newpage
\begin{figure}
\setlength{\unitlength}{0.000098\columnwidth}
\begin{picture}(0,0)%
\psfig{file=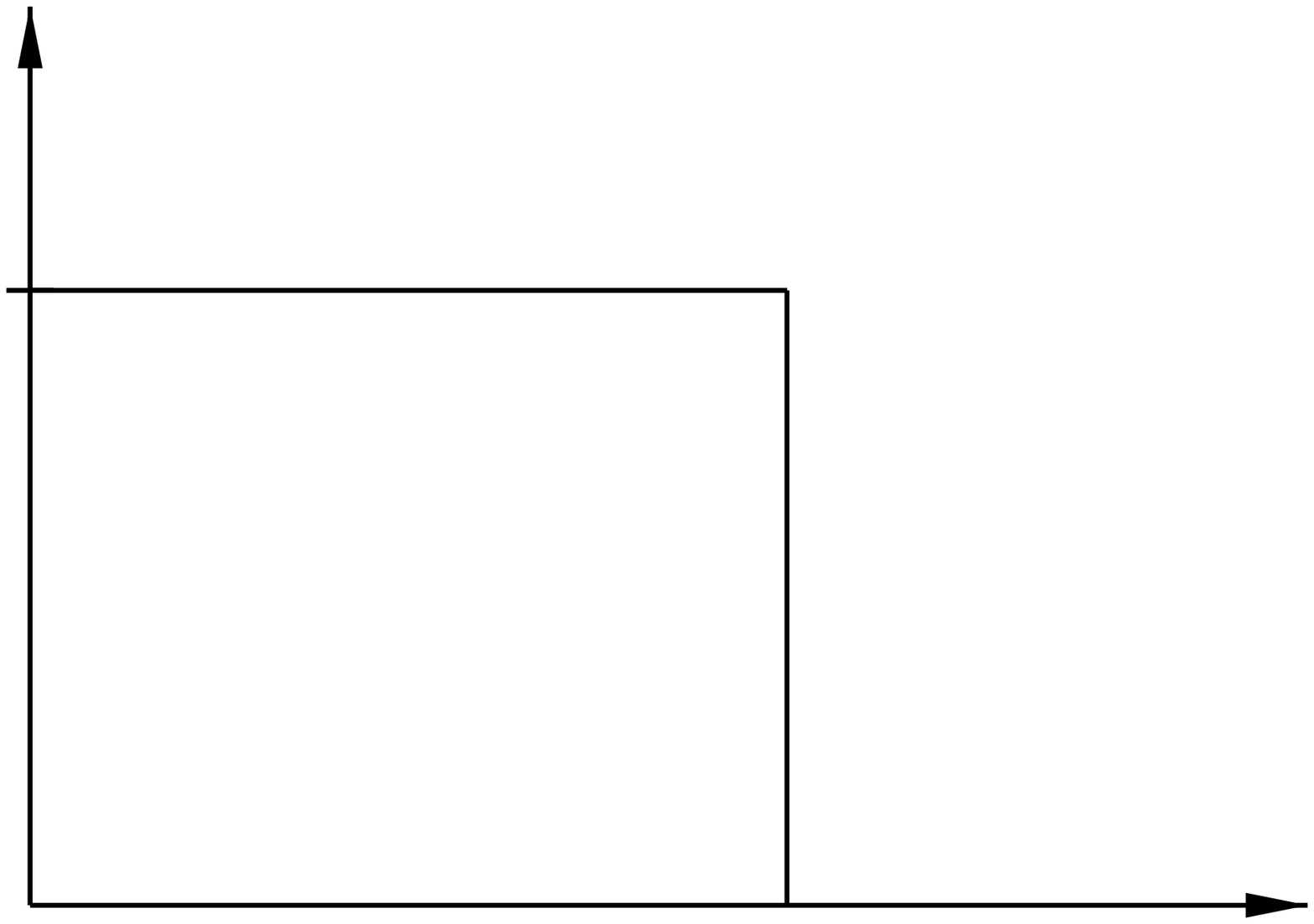,width=0.85\columnwidth}%
\end{picture}%
\begin{picture}(9108,6195)(1426,-8311)
\put(1951,-4036){\makebox(0,0)[lb]{1}}
\put(4501,-6061){\makebox(0,0)[lb]{BEC}}
\put(3501,-3561){\makebox(0,0)[lb]{BS (Bose-Solid)}}
\put(10001,-7411){\makebox(0,0)[lb]{$\frac{|V|}{t}$}}
\put(1426,-2311){\makebox(0,0)[lb]{$n_B$}}
\put(8126,-6061){\makebox(0,0)[lb]{phase separation}}
\put(7176,-8311){\makebox(0,0)[lb]{2}}
\end{picture}
\vskip 0.5cm
\caption{Qualitative phase diagram of the 2D Bose gas with the Van der Waals interaction on the
square lattice.}
\label{fig:2}
\end{figure}

\begin{thebibliography}{99}
\bibitem{Valatin58} J.G. Valatin and D. Butler, Nuovo Cimento { \bf 10}, 37 (1958).
\bibitem{Nozieres82}  P. Nozieres and D. Saint James, J. Phys. (Paris) {\bf 43}, 1133
(1982) and references therein.
\bibitem{Jordanski65} S.V.Jordanski, Zh. Eksp. Teor. Fiz.{\bf 47}, 167 (1964) 
[Sov.Phys.JETP, {\bf
20}, 112 (1965)].
\bibitem{Rice88} M.J.~Rice and Y.R.~Wang, Phys. Rev. B {\bf 37}, 5893 (1988).
\bibitem{Emery90} V.J.~Emery, S.A.~Kivelson, and H.Q.~Lin, Phys. Rev. Lett. {\bf 64},
475 (1990).
\bibitem{Kagan94} M.Yu.~Kagan and T.M.~Rice, J.Phys.: Condens. Matt. {\bf 6},
3771 (1994).
\bibitem{Fisher89}M.P.A.~Fisher, P.B.~Weichman,  G.~Grinstein, and D.S.~Fisher,
Phys. Rev. B {\bf 40}, 546  (1989);
P.B.~Weichman, M.~Rasolt, M.E.~Fisher, and M.J.Stephen, Phys. Rev. B {\bf 33},
4632, 1986.
\bibitem{Monien94} J.K.~Freezicks and H.Monien, Europhys. Lett. {\bf 26}, 545
(1994).
\bibitem{Kotliar91} D.S.~Rokhsar and B.G. Kotliar, Phys. Rev. B {\bf 44}, 10328
(1991).
\bibitem{Kampf93} A.P.~Kampf and G.T.~Zimanyi, Phys. Rev. B {\bf 47}, 279
(1993) ; K.Sheshardi et al, Europhys. Lett. {\bf 22}, 257 (1993).
\bibitem{Griffin98} H.~Shi and A.~Griffin, Phys. Rep. {\bf 304}, 1 (1998).
\bibitem{Miyake83} K.~Miyake, Prog. Theor. Phys {\bf 69}, 1794 (1983).
\bibitem{Alexandrov81} A.~Alexandrov, J.~Ranninger,
Phys. Rev. B {\bf 23}, 1796 (1981).
\bibitem{Nozieres85} P.~Nozieres, S.Schmitt-Rink, J. Low Temp. Phys. {\bf 59},
195 (1985).
\bibitem{Kagan00} M.Yu.~Kagan, R.Fresard, M.~Capezzali, and H.~Beck,
Phys. Rev. B {\bf 57}, 5995 (1998),
Physica B {\bf 284-288}, 447 (2000).
\bibitem{Fisher88} D.S.~Fisher and P.C.~Hohenberg, Phys. Rev. B{\bf 37},
4936 (1988).
\bibitem{Popov72} V.N.~Popov, Theor. Math. Phys. {\bf 11}, 565 (1972).
\bibitem{Kosterlitz73} J.M.~Kosterlitz and D.J.~Thouless, J of Phys. C {\bf 6},
1181 (1973); V.L.~Berezsinski,  Zh. Eksp. Teor. Fiz. { \bf 61}, 610 (1972),
[Sov. Phys. JETP  {\bf 34}, 610 (1972)].
\bibitem{Dagotto93} E.~Dagotto and J.~Riera, Phys. Rev. Lett. {\bf 70}, 682
(1993); E.~Dagotto, J.~Riera et al, Phys.Rev.B {\bf 49}, 3548 (1994), Appendix B.
\bibitem{Nozieres98} G.~R\"opke, A. Schnell, P.~Schuck, and P.Nozieres,
Phys. Rev. Lett {\bf 80}, 3177 (1998).
\bibitem{Keldysh92} L.V.~Keldysh, Solid State Commun. {\bf 84}, 37 (1992),
V.S.~Babichenko,  Zh. Eksp. Teor. Fiz. {\bf 64}, 612 (1973)[ Sov. Phys. JETP {\bf 37}, 311 (1973)].
\bibitem{Lozovik99} Yu.E.Lozovik, O.L.Berman, M.Villander,
Zh. Eksp. Teor. Fiz. {\bf 115}, 1786 (1999) [Sov.Phys.JETP, {\bf 88}, 980 (1999)].
\bibitem{Anderson87} P.W.~Anderson, Science {\bf 235}, 1196 (1987).
\bibitem{Wen98} X.G.~Wen and P.A.~Lee, Phys. Rev Lett. {\bf 80}, 2193  (1998).
 \bibitem{Kotliar88} G.Kotliar and J.~Liu, Phys. Rev. B {\bf 38}, 5142  (1988).
\bibitem{Lee98}  P.A.~Lee, N.~Nagaosa, T.-K.~Ng, X.G.~Wen, Phys. Rev. B {\bf
57}, 6003 (1998).
\bibitem{Salk00}  S.-S.~Lee, S.H.S.~Salk, cond-mat/0001218 (2000).
\bibitem{Lee92} P.A.~Lee, N.~Nagaosa, Phys. Rev. B {\bf 46}, 5621 (1992).
\bibitem{Arovas88} D.P.~Arovas, A.Auerbach, Phys.Rev.B {\bf
38}, 316 (1988); A.~Chubukov, Phys.Rev.B {\bf 44}, 12318 (1991).
\bibitem{Irhin99} V.Yu.Irkhin, A.A.Katanin, M.I.Katsnelson,
Phys.Rev.B {\bf 60}, 1082 (1999).
 \end{thebibliography}
\end{document}